\documentclass[12pt]{article}
\usepackage{graphicx}
\usepackage{mathrsfs}
\usepackage[utf8]{inputenc} 
\usepackage{tikz}
\usepackage{graphicx} 
\usepackage{booktabs} 
\usepackage{array} 
\usepackage{paralist} 
\usepackage{verbatim} 
\usepackage{subfig} 
\usepackage{amssymb} 
\usepackage{amsmath} 
\usepackage{multirow} 
\usepackage{rotating} 
\usepackage[version=3]{mhchem}  
\usepackage[affil-it]{authblk}
\usepackage{mathrsfs}
\usepackage{mathabx}
\usepackage{fancyhdr} 
\usepackage{longtable}
\newcommand{\bi}[1]{\bf{#1}}
\begin{document}
\title{One dimensional zone center phonons in \(O_{h}\) space group}
\author{Jian Li%
  \thanks{\texttt{jianli@sci.ccny.cuny.edu}}, 
Jiufeng J. Tu, Joseph L. Birman}
\affil{Physics Department, The City College of New York,\\
160 Convent Avenue, New York 10031, USA}

\maketitle

\begin{abstract}
Motivated by experiments, we performed a systematic study on the one dimensional zone center phonons in \(O_{h}\) space groups. All the one dimensional phonon modes for different Wyckoff positions are tabulated. We show that at least four (inequivalent) atoms (in one set of Wyckoff positions) are needed to carry a single one dimensional phonon. A general restriction rule on the number of atoms and the number of one dimensional phonons is obtained. The same restriction applies to phonons of all cubic crystal systems (\(T\), \(T_{h}\), \(T_{d}\), \(O\), \(O_{h}\)) and to magnons for crystals whose unitary symmetry elements are in (\(T\), \(T_{h}\), \(T_{d}\), \(O\), \(O_{h}\)). Crystals with \(A15\) structure are found to satisfy experimental requirements (to have a \(\Gamma_{2}^{+}\) phonon) while most crystals of \(O_{h}\) space groups do not. \(A15\) is also used to demonstrate the rules we found regarding the phonon strucutre in \(O_{h}\) space groups.
\end{abstract}


\section{Introduction}

In doing spectroscopy, one often needs certain phonon modes to test selection rules. In our lab, a novel Raman scattering experiment is being conducted and \(\Gamma_{2}^{+}\) phonon of the \(O_{h}\) space group is among the several modes to produce new features. However, seaching for \(\Gamma_{2}^{+}\) phonon in \(O_{h}\) space groups was not successful. Many well known crystals in \(O_{h}\) space group, such as silicon and perovskite, do not have \(\Gamma_{2}^{+}\) phonons. A careful examination of those crystals shows that not only \(\Gamma_{2}^{+}\), but also all the one ( \(\Gamma_{1}^{+}\), \(\Gamma_{2}^{+}\), \(\Gamma_{1}^{-}\), \(\Gamma_{2}^{-}\)) and two ( \(\Gamma_{3}^{+}\), \(\Gamma_{3}^{-}\)) dimensional phonons rarely happen while three dimensional phonons (\(\Gamma_{4}^{+}\), \(\Gamma_{5}^{+}\), \(\Gamma_{4}^{-}\), \(\Gamma_{5}^{-}\)) always present.

Instead of checking phonon structures of available crystals one by one, we perform a systematic study on the zone center phonon structures of \(O_{h}\) space groups. All one dimensional phonons are listed for different Wyckoff positions of the ten \(O_{h}\) space groups. Analysis shows that at least four inequivalent atoms in one set of Wyckoff positions are required to have one dimensional phonons. This explains the absence of the \(\Gamma_{2}^{+}\) phonons in NaCl, diamond and other crystals with simple structures. The results are tabulated and, with the help of our tables, one can choose proper crystals when a certain one or two dimensional phonon is needed. As far as we know, there are no similar analysis in the literature. The fact that crystals belonging to \(O_{h}\) space groups are the most common ones~\cite{crystaldata} makes our work useful. A restriction relation between number of atoms in one set of Wyckoff position and the number of one dimensional phonons (magnons) is found in cubic lattice systems (\(T\), \(T_{h}\), \(T_{d}\), \(O\), \(O_{h}\)). All symmetry assignments follow the Koster notations~\cite{KDWS}.

This note is arranged in the following order: Section II introduces the method to obtain one dimensional zone center phonons; The results are tabulated in section III for all ten \(O_{h}\) space groups at all Wyckoff positions; Section IV explores higher dimensional phonons: a certain number of two and three dimensional phonons accompany a single one dimensional phonon. The restriction relation on the number of one dimensional phonons and the number of atoms in one Wyckoff set is presented; Section V applies the same argument to all cubic lattice systems (\(T\), \(T_{h}\), \(T_{d}\), \(O\), \(O_{h}\)) and the same restriction is found; We discuss the structures of magnons in section VI where the same restrictions are found; Discussions are in section VII, where the phonon structure of \(A15\) crystals is studied.

\section{Theoretical background}

\(O_{h}\) space groups are space groups with \(O_{h}\) point group. They are \(O_{h}^{1}\) (\(Pm3m\)), \(O_{h}^{2}\) (\(Pn3n\)), \(O_{h}^{3}\) (\(Pm3n\)), \(O_{h}^{4}\) (\(Pn3m\)), \(O_{h}^{5}\) (\(Fm3m\)), \(O_{h}^{6}\) (\(Fm3c\)), \(O_{h}^{7}\) (\(Fd3m\)), \(O_{h}^{8}\) (\(Fd3c\)), \(O_{h}^{9}\) (\(Im3m\)) and \(O_{h}^{10}\) (\(Ia3d\)). In those space groups, 48 rotational operations are associated with simple cubic, face-centered-cubic or body-centered-cubic lattices. The general form of a symmetry operation is \{\(\hat g|\bi{R_{L}} + \bi{\tau_{g}}\)\}: \(\hat{g}\) is the rotational operation, \(\bi{R_{L}}\) is the lattice translation. \(\bi{\tau_{g}}\) is 0 for symmorphic space groups and some fractional translation(s) for nonsymmorphic space groups.

\begin{table}
\caption{The character table of \(O_{h}\) point group.and the direct product of \(\Gamma_{4}^{-}\) with all representations of \(O_{h}\) point group~\cite{KDWS}.} \label{tab:charactertable}
\begin{center}
\begin{tabular}{|l|r|r|r|r|r|r|r|r|r|r|c|}\hline
& \(E\) & 8\(C_{3}\) &3\(C_{2}\) & 6\(C_{4}\) & 6\(C_{2}\) & \(I\) & 8\(S_{6}\) & 3\(\sigma_{h}\) & 3\(S_{4}\) & 6\(\sigma_{d}\)&\(\otimes\) \(\Gamma_{4}^{-}\) \\ \hline
\(\Gamma_{1}^{+}\) & 1 & 1 & 1 & 1 & 1 & 1 & 1 & 1 & 1 & 1 &\(\Gamma_{4}^{-}\)\\ \hline
\(\Gamma_{2}^{+}\) & 1 & 1 & 1 & -1 & -1 & 1 & 1 & 1 & -1 & -1 &\(\Gamma_{5}^{-}\)\\ \hline
\(\Gamma_{3}^{+}\) & 2 & -1 & 2 & 0 & 0 & 2 & -1 & 2 & 0 & 0&\(\Gamma_{4}^{-}\) + \(\Gamma_{5}^{-}\)\\ \hline
\(\Gamma_{4}^{+}\) & 3 & 0 & -1 & 1 & -1 & 3 & 0 & -1 & 1 & -1 &\(\Gamma_{1}^{-}\)+\(\Gamma_{3}^{-}\)+\(\Gamma_{4}^{-}\)+\(\Gamma_{5}^{-}\) \\ \hline
\(\Gamma_{5}^{+}\) & 3 & 0 & -1 & -1 & 1 & 3 & 0 & -1 & -1 & 1& \(\Gamma_{2}^{-}\)+\(\Gamma_{3}^{-}\)+\(\Gamma_{4}^{-}\)+\(\Gamma_{5}^{-}\) \\ \hline
\(\Gamma_{1}^{-}\) & 1 & 1 & 1 & 1 & 1 & -1 & -1 & -1 & -1 & -1 &\(\Gamma_{4}^{+}\)\\ \hline
\(\Gamma_{2}^{-}\) & 1 & 1 & 1 & -1 & -1 & -1 & -1 & -1 & 1 & 1 &\(\Gamma_{5}^{+}\)\\ \hline
\(\Gamma_{3}^{-}\) & 2 & -1 & 2 & 0 & 0 & -2 & 1 & -2 & 0 & 0 &\(\Gamma_{4}^{+}\) + \(\Gamma_{5}^{+}\)\\ \hline
\(\Gamma_{4}^{-}\) & 3 & 0 & -1 & 1 & -1 & -3 & 0 & 1 & -1 & 1& \(\Gamma_{1}^{+}\)+\(\Gamma_{3}^{+}\)+\(\Gamma_{4}^{+}\)+\(\Gamma_{5}^{+}\) \\ \hline
\(\Gamma_{5}^{-}\) & 3 & 0 & -1 & -1 & 1 & -3 & 0 & 1 & 1 & -1& \(\Gamma_{2}^{+}\)+\(\Gamma_{3}^{+}\)+\(\Gamma_{4}^{+}\)+\(\Gamma_{5}^{+}\) \\ \hline
\end{tabular}
\end{center}
\end{table}

We now proceed to determine the phonon modes. Let \(\gamma\) denote any of the one dimensional phonons (\(\Gamma_{1}^{+}, \Gamma_{2}^{+}, \Gamma_{1}^{-}, \Gamma_{2}^{-}\)). \(\gamma\) is carried by a set of \(n\) atoms labelled by their Wyckoff positions, such as \(12(e)\) in \(O_{h}^{2}\). Among the 48 rotational operations, \(48/n\) of them, denoted by \{\(\hat \alpha|\bi{\tau_{\alpha}}\)\}, do not move the atom or move the atom to its equvalent position: \( \{\hat \alpha | \bi\tau_{\alpha}\}  \bi{r_i}=\bi{r_j}\) and \(\bi{r_j} = \bi{r_i} +\bi{R_{L}}\) where\(\bi{r_i}\) is the position of the original atom and \(\bi{R_{L}}\) is any lattice vector; the rest (\(48-48/n\)) operations, denoted by \(\{\hat \beta | \bi{\tau_{\beta}}\}\), would shift the original atom to its inequivalent positions: \(\{ \hat \beta | \bi{\tau_{\beta}} \} \bi{r_i}=\bi{r_j}\) and \(\bi{r_j} \neq \bi{r_i} + \bi{R_{L}}\). If a set of \(n\) atoms carry \(\gamma\) phonon, the atomic displacement vector \(\bi{V}\) at atom \(\bi{r_{i}}\) must satisfy: \(\hat {g} \cdot \bi{V}  =  \chi^{\gamma}(\hat {g}) \cdot \bi{V} \), where \(\hat g\) is any of the 48 rotational operations. In general, \(\bi{V}\) is in the form of (\(v_{x}, v_{y}, v_{z}\)) and any solution of \(\bi{V} = (v_{x}, v_{y}, v_{z})\) gives the allowed phonon mode. With the defination of \(\hat \alpha\) the set of 48 equations can be simplified: \(\hat \alpha \cdot \bi{V}  =  \chi^{\gamma}(\hat {\alpha}) \cdot \bi{V}\), which reduce the number of equations down to \(48/n\). That is, only operations that shift the atom to itself or its equivalent positions need to be considered. The existance of solutions for \(\bi{V}\) is necessary and sufficient for the existance of one dimensional phonons.

\section{Results on one dimensional phonons}

All Wyckoff positions in \(O_{h}\) space groups are considered in this section. When dealing with phonons, the primitive cell is more suitable than the unit cell. Therefore, a slightly different notation is adopted, compared to the {\em International Table for Crystallography}~\cite{Wyckoff}: although the letters of Wyckoff positions remain the same, we put down the number of atoms in one primitive cell instead of one unit cell. For example, in \(O_{h}^{10}\), \(48(h)\) in this note is the \(96(h)\) in the {\em International Table for Crystallography}. Directions of phonon modes for (\(\Gamma_{1}^{+}\), \(\Gamma_{2}^{+}\), \(\Gamma_{1}^{-}\), \(\Gamma_{2}^{-}\)) phonons are listed in table~\ref{tab:Oh}.

\begin{center}
\begin{longtable}{|l|l|l|l|l|} \caption{One dimentional phonon modes in \(O_{h}\) space groups. \label{tab:Oh}}\\
\hline
Wyckoff positions & \(\Gamma_{1}^{+}\) & \(\Gamma_{2}^{+}\) & \(\Gamma_{1}^{-}\) & \(\Gamma_{2}^{-}\) \endhead
\hline \endlastfoot
\hline 
\multicolumn{5}{l}{\(O_{h}^{1}\)} \\ \hline
48, (n), (x, y, z)&(x, y, z)&(x, y, z)&(x, y, z)&(x, y, z) \\ \hline
24, (m), (x, x, z)&(x, x, z)&(x, -x, 0)&(x, -x, 0)&(x, x, z) \\  \hline
24, (l), (1/2, y, z)&(0, y, z)&(0, y, z)&(x, 0, 0)&(x, 0, 0) \\ \hline
24, (k), (0, y, z)&(0, y, z)&(0, y, z)&(x, 0, 0)&(x, 0, 0) \\ \hline
12, (j), (1/2, y, y)&(0, y, y)&(0, y, -y)&&(x, 0, 0) \\ \hline
12, (i), (0, y, y)&(0, y, y)&(0, y, -y)	&&(x, 0, 0) \\ \hline
12, (h), (x, 1/2, 0)&(x, 0, 0)&(x, 0, 0)&& \\ \hline
 8, (g), (x, x, x)&(x, x, x)&&&(x, x, x)\\ \hline
 6, (f), (x, 1/2, 1/2)&(x, 0, 0)&&&\\	 \hline
 6, (e), (x, 0, 0)&(x, 0, 0)&&&\\ \hline
 3, (d), (1/2, 0, 0)& & & &\\	 \hline
 3, (c), (0, 1/2, 1/2)& & & &\\	 \hline	
 1, (b), (1/2, 1/2, 1/2)& & & &\\	 \hline	
 1, (a), (0, 0, 0)&&&&\\	 \hline
\multicolumn{5}{l}{\(O_{h}^{2}\)}  \\ \hline
48, (i), (x, y, z)&(x, y, z)&	(x, y, z)&(x, y, z)&(x, y, z) \\ \hline
24, (h), (0, y, y)&(0, y, y)&(x, y, -y)&(0, y, y)&(x, y, -y) \\ \hline
24, (g), (x, 0, 1/2)&(x, 0, 0)&(x, 0, 0)&(x, 0, 0)&(x, 0, 0) \\ \hline
16, (f), (x, x, x)&(x, x, x)&(x, x, x)&(x, x, x)&(x, x, x) \\ \hline
12, (e), (x, 0, 0)&(x, 0, 0)&&(x, 0, 0)& \\ \hline
12, (d), (1/4, 0, 1/2)&&(x, 0, 0)&&(x, 0, 0) \\ \hline
8, (c), (1/4, 1/4, 1/4)&&&(x, x, x)&(x, x, x) \\ \hline
6, (b), (0, 1/2, 1/2)&&&&	\\ \hline		
2, (a), (0, 0, 0)&&&&	\\ \hline	
\multicolumn{5}{l}{\(O_{h}^{3}\)} \\ \hline
48, (l), (x, y, z)&(x, y, z)&(x, y, z)&(x, y, z)&(x, y, z) \\ \hline
24, (k), (0, y, z)&(0, y, z)&(0, y, z)&(x, 0, 0)&(x, 0, 0) \\ \hline
24, (j), (1/4, y, y+1/2)&(0, y, y)&(x, y, -y)&(0, y, y)&(x, y, -y) \\ \hline
16, (i), (x, x, x)&(x, x, x)&(x, x, x)&(x, x, x)&(x, x, x) \\ \hline
12, (h), (x, 1/2, 0)&(x, 0, 0)&(x, 0, 0)&&\\ \hline
12, (g), (x, 0, 1/2)&(x, 0, 0)&(x, 0, 0)&&\\ \hline
12, (f), (x, 0, 0)&(x, 0, 0)&(x, 0, 0)&&\\ \hline
8, (e), (1/2, 1/2, 1/2)&&(x, x, x)&&(x, x, x)\\ \hline
6, (d), (1/4, 1/2, 0)&&(x, 0, 0)&&\\ \hline
6, (c), (1/4, 0, 1/2)&&(x, 0, 0)&&\\ \hline
6, (b), (0, 1/2, 1/2)&&&&	\\ \hline
2, (a), (0, 0, 0)&&&&	\\ \hline
\multicolumn{5}{l}{\(O_{h}^{4}\)} \\ \hline
48, (l), (x, y, z)&(x, y, z)&(x, y, z)&(x, y, z)&(x, y, z) \\ \hline
24, (k), (x, x, z)&(x, x, z)&(x, -x, 0)&(x, -x, 0)&(x, x, z) \\ \hline
24, (j), (1/4, y, y+1/2)&(0, y, y)&(x, y, -y)&(0, y, y)&(x, y, -y) \\ \hline
24, (i), (1/4, y, -y+1/2)&(0, y, -y)&(x, y, y)&(0, y, -y)&(x, y, y) \\ \hline
24, (h), (x, 0, 1/2)&(x, 0, 0)&(x, 0, 0)&(x, 0, 0)&(x, 0, 0) \\ \hline
12, (g), (x, 0, 0)&(x, 0, 0)&&&(x, 0, 0) \\ \hline
12, (f), (1/4, 0, 1/2)&&(x, 0, 0)&&(x, 0, 0) \\ \hline
8, (e), (x, x, x)&(x, x, x)&&&(x, x, x)\\ \hline
6, (d), (0, 1/2, 1/2)	&&&& \\ \hline
4, (c), (3/4, 3/4, 3/4)&&&&(x, x, x) \\ \hline
4, (b), (1/4, 1/4, 1/4)&&&&(x, x, x) \\ \hline
2, (a), (0, 0, 0)&&&&\\ \hline
\multicolumn{5}{l}{\(O_{h}^{5}\)} \\ \hline
48, (l), (x, y, z)&	(x, y, z)&(x, y, z)&(x, y, z)&(x, y, z)\\ \hline
24, (k), (x, x, z)&(x, x, z)&(x, -x, 0)&(x, -x, 0)&(x, x, z)\\ \hline
24, (j), (0, y, z)&	(0, y, z)&(0, y, z)&(x, 0, 0)&(x, 0, 0)\\ \hline
12, (i), (1/2, y, y)&(0, y, y)&(0, y, -y)&&(x, 0, 0)\\ \hline
12, (h), (0, y, y)&(0, y, y)&(0, y, -y)&&(x, 0, 0)\\ \hline
12, (g), (x, 1/4, 1/4)&(x, 0, 0)&&&(x, 0, 0)\\ \hline
8, (f), (x, x, x)&(x, x, x)&&&(x, x, x)\\ \hline
6, (e), (x, 0, 0)&(x, 0, 0)&&&\\ \hline
6, (d), (0, 1/4, 1/4)&&&&(x, 0, 0)\\ \hline
2, (c), (1/4, 1/4, 1/4)&&&&\\ \hline
1, (b), (1/2, 1/2, 1/2)&&&&\\ \hline
1, (a), (0, 0, 0)&&&&\\ \hline
\multicolumn{5}{l}{\(O_{h}^{6}\)} \\ \hline
48, (j), (x, y, z)&(x, y, z)&(x, y, z)&(x, y, z)&(x, y, z)\\ \hline
24, (i), (0, y, z)&(0, y, z)&(0, y, y)&(x, 0, 0)&(x, 0, 0)\\ \hline
24, (h), (1/4, y, y)&(0, y, y)&(x, y, -y)&	(0, y, y)&(x, y, -y)\\ \hline
16, (g), (x, x, x)&(x, x, x)&(x, x, x)&(x, x, x)&(x, x, x)\\ \hline
12, (f), (x, 1/4, 1/4)&(x, 0, 0)&(x, 0, 0)&&\\ \hline
12, (e), (x, 0, 0)&(x, 0, 0)&(x, 0, 0)&&\\ \hline
6, (d), (0, 1/4, 1/4)&&&(x, 0, 0)&\\ \hline
6, (c), (1/4, 0, 0)&&(x, 0, 0)&&\\ \hline	
2, (b), (0, 0, 0)&&&&	\\ \hline
2, (a), (1/4, 1/4, 1/4)&&&&\\ \hline
\multicolumn{5}{l}{\(O_{h}^{7}\)} \\ \hline
48, (i), (x, y, z)&(x, y, z)&(x, y, z)&(x, y, z)&(x, y, z)\\ \hline
24, (h), (1/8, y, -y+1/4)&(0, y, -y)&(x, y, y)&(0, y, -y)&(x, y, y)\\ \hline
24, (g), (x, x, z)&(x, x, z)&(x, -x, 0)&(x, -x, 0)&	(x, x, z)\\ \hline
12, (f), (x, 0, 0)&(x, 0, 0)	&&&(x, 0, 0)\\ \hline
8, (e), (x, x, x)&(x, x, x)&&&(x, x, x)\\ \hline
4, (d), (5/8, 5/8, 5/8)&&&&(x, x, x)\\ \hline
4, (c), (1/8, 1/8, 1/8)&&&&(x, x, x)\\ \hline
2, (b), (1/2, 1/2, 1/2)&&&&\\ \hline
2, (a), (0, 0, 0)&&&&	\\ \hline
\multicolumn{5}{l}{\(O_{h}^{8}\)} \\ \hline
48, (h), (x, y, z)&(x, y, z)&(x, y, z)&(x, y, z)	&(x, y, z)\\ \hline
24, (g), (1/8, y, -y+1/4)&(0, y, -y)&(x, y, y)&(0, y, -y)&(x, y, y)\\ \hline
24, (f), (x, 0, 0)&(x, 0, 0)&(x, 0, 0)&(x, 0, 0)&(x, 0, 0)\\ \hline
16, (e), (x, x, x)&(x, x, x)&(x, x, x)&(x, x, x)&(x, x, x)\\ \hline
12, (d), (1/4, 0, 0)&&(x, 0, 0)&(x, 0, 0)&\\ \hline
8, (c), (3/8, 3/8, 3/8)&&&(x, x, x)&(x, x, x)\\ \hline
8, (b), (1/8, 1/8, 1/8)&&	(x, x, x)&&(x, x, x)\\ \hline
4, (a), (0, 0, 0)&&&&	\\ \hline
\multicolumn{5}{l}{\(O_{h}^{9}\)} \\ \hline
48, (l), (x, y, z)&(x, y, z)&	(x, y, z)&(x, y, z)&(x, y, z)\\ \hline
24, (k), (x, x, z)&(x, x, z)&(x, -x, 0)&(x, -x, 0)&(x, x, z)\\ \hline
24, (j), (0, y, z)&(0, y, z)&(0, y, y)&(x, 0, 0)&(x, 0, 0)\\ \hline
24, (i), (1/4, y, -y+1/2)&(0, y, -y)&(x, y, y)&(0, y, -y)&(x, y, y)\\ \hline
12, (h), (0, y, y)&(0, y, y)&(0, y, -y)&&(x, 0, 0)\\ \hline
12, (g), (x, 0, 1/2)&(x, 0, 0)&(x, 0, 0)&&\\ \hline
8, (f), (x, x, x)&(x, x, x)&&&(x, x, x)\\ \hline
6, (e), (x, 0, 0)&(x, 0, 0)&&&\\ \hline
6, (d), (1/4, 0, 1/2)&&(x, 0, 0)&&\\ \hline
4, (c), (1/4, 1/4, 1/4)&&&&(x, x, x)\\ \hline
3, (b), (0, 1/2, 1/2)&&&&\\ \hline
1, (a), (0, 0, 0)&&&&\\ \hline
\multicolumn{5}{l}{\(O_{h}^{10}\)} \\ \hline
48, (h), (x, y, z)&(x, y, z)&(x, y, z)& (x, y, z)&(x, y, z) \\ \hline
24, (g), (1/8, y, -y+1/4)&	(0, y, -y)&(x, y, y)&(0, y, -y)&(x, y, y) \\ \hline
24, (f), (x, 0, 1/4)&(x, 0, 0)&(x, 0, 0)&(x, 0, 0)&(x, 0, 0) \\ \hline
16, (e), (x, x, x)&(x, x, x)&(x, x, x)&(x, x, x)&(x, x, x) \\ \hline
12, (d), (3/8, 0, 1/4)&&(x, 0, 0)&(x, 0, 0)& \\ \hline
12, (c), (1/8, 0, 1/4)&&(x, 0, 0) &&(x, 0, 0) \\ \hline
8, (b), (1/8, 1/8, 1/8)&&(x, x, x)&&(x, x, x) \\ \hline
8, (a), (0, 0, 0)&&&(x, x, x)&(x, x, x) \\ \hline
\end{longtable}
\end{center}

The meaning of the notations in table~\ref{tab:Oh} is as follows: for a certain phonon mode belonging to a set of atoms, the number of free parameters in the phonon modes is the number of its appearances. For example, in \(O_{h}^{10}\), \(\Gamma_{2}^{+}\) phonon of Wyckoff position \(24(g)\) \((1/8, y, -y+1/4)\) is labelled \((x, y, y)\). It means that \(\Gamma_{2}^{+}\) phonon modes appear twice and the \(\Gamma_{2}^{+}\) phonon modes on atom \((1/8, y, -y+1/4)\) are in the (1, 0, 0) direction and (0, 1, 1) direction. The actual atomic displacements are the linear combinations of the two phonon modes, the coefficients of which cannot be determined by symmetry alone. It is worth mentioning that the number of \(\Gamma_{1}^{+}\) phonons equals the number of free parameters of the Wyckoff positions because totally symmetric distortions preserve the symmetry. For example, Wyckoff position \(24(g)\) \((1/8, y, -y+1/4)\) in \(O_{h}^{10}\) has only one \(\Gamma_{1}^{+}\) phonon mode which is in the (0, 1, -1) direction. Also one can see that any one dimensional phonon does not happen more than three times in one set of Wyckoff positions.

\section{Two and three dimensional phonons}

Additional informations about two and three dimensional phonons can also be obtained from the analysis on one dimensional phonons. In general, the phonon structure is determined by the decomposition of mechanical representation, which is the direct product of permutation group and the vector representation~\cite{Birman}. Permutation group is a group formed by taking atomic positions as basis functions. Its character under a certain symmetry operation equals the number of atoms unchanged or can be shifted back to itself by lattice vector. The vector representation is formed with basis functions (\(x, y, z\)) and it belongs to \(\Gamma_{4}^{-}\) in \(O_{h}\) space groups. The direct products of \(\Gamma_{4}^{-}\) with different representations are given in table~\ref{tab:charactertable}.

Careful inspection of the tables shows many interesting results. There will always be three dimensional phonons, no matter how simple the structure is. Each one dimensional phonon is accompanied by one two dimensional phonon and two three dimensional phonons. This makes a restriction on the number of appearances of one dimensional phonons: \(n \times 3 \geq \sum_{\gamma}{m \times 9} \label{eq:restriction1}\), where \(n\) is the number of atoms in one set of Wyckoff position, and \(m\) is number of one dimensional phonons. Summation over \(\gamma\) means summation over all the one dimensional phonons.

The character of the permutation group must be non-negative. One dimensional phonons correspond to \(\Gamma_{4}^{+}\), \(\Gamma_{5}^{+}\), \(\Gamma_{4}^{-}\), \(\Gamma_{5}^{-}\) in the permutation representation and they all have negative characters under \(C_{2}\) symmetry operation (see table~\ref{tab:charactertable}). Other one or two dimensional representations which have positive characters under \(C_{2}\) operation must present in the permutation representation. It makes further restrictions on the number of appearances of one dimensional phonons:
\begin{equation}
n \times 3 \geq \sum_{\gamma}{m \times 12} \label{eq:restriction2}
\end{equation}
This means that at least four atoms are needed to carry a single one dimensional phonon, and this rule is indeed observed in all ten \(O_{h}\) space groups (table~\ref{tab:Oh}). The difference between \( {\sum_{\gamma} {m \times 12}}\) and \(m \times 3\) will be totally filled by three dimensional phonons.

\section{Cubic lattice system}

The same argument can be applied to other cubic lattice systems: \(T\), \(T_{h}\), \(T_{d}\) and \(O\). The direct product tables of those point groups \cite{KDWS} show that any one dimensional phonon is accompanied by one two dimensional phonon and two three dimensional phonons, giving a total dimension of 9. Those 9 dimensional phonons correspond to one three dimensional representation in the permutation representation, and all three dimensional representations have negative characters under one set of \(C_{2}\) operations (there are two sets of \(C_{2}\) operations in \(O\) and \(O_{h}\)). Permutation representation must have non-negative characters, therefore other one or two dimensional representations with positive characters under \(C_{2}\) must present. This makes the restriction for \(T\), \(T_{h}\), \(T_{d}\), \(O\) and \(O_{h}\) space groups: \(n \times 3 \geq \sum_{\gamma}{m \times 12} \label{eq:restriction3}\) and the summation over \(\gamma\) is the summation over all one dimensional phonons in \(T\), \(T_{h}\), \(T_{d}\), \(O\) and \(O_{h}\) space groups. It should be noticed that, for \(T\) and \(T_{h}\), the two dimensional representation is actually two one dimensional representations forced to stick together due to time reveral symmetry.

The same analysis on one dimensional phonons that lead to table~\ref{tab:Oh} can be extended to \(T\), \(T_{h}\), \(T_{d}\) and \(O\) space groups. However, due to limitation of space, they are not tabulated.

\section{Magnons for magnetic space groups with cubic unitary group}

Without external magnetic field, time reversal operator \(\theta\) is also a symmetry operation of the system. The inclusion of time reversal operator generates magnetic space group \(\bf M\) which contains equal number of unitary and antiunitary elements: \(\bf M\) = \(\bf H\) + \(\bf A \bf H\). \(\bf H\) is ordinary space group and \(\bf A\) is antiunitary coset representative. The magnon symmetry is characterized within space group \(\bf H\)~\cite{dimmock}. Representations of magnons are contained in the direct product of permutation group and the ``pseudo vector representation" (\(R_{x}\), \(R_{y}\), \(R_{z}\)).

For those magnetic space groups with \(\bf H\) being one of the cubic space groups (\#195 -- \#230), we find that the same analysis on phonons can also be applied. ``Pseudo vector" belongs to one of the three dimensional representations in \(T\), \(T_{h}\), \(T_{d}\), \(O\) and \(O_{h}\) point groups and any one or two dimensional representations must be contained in the direct product of ``pseudo vector group" and a three dimensional representation of the permutation group. As for phonons, these three dimensional representations have negative character under \(C_{2}\) operation therefore some one or two dimensional representations must present in the permutation group whose characters must be non-negative. This leads to the same restrictions that at least four atoms are needed to have one or two dimensional magnons if the unitary group \(\bf H\) of the magnetic space group \(\bf M\) belongs to cubic space groups (\#195 -- \#230).

\section{Discussion}

We are now in position to solve the problem that initiated this study. Table~\ref{tab:Oh} shows that the \(\Gamma_{2}^{+}\) requires 6 inequivalent atoms for \(O_{h}^{3}\),  \(O_{h}^{6}\), \(O_{h}^{9}\), 8 inequivalent atoms for \(O_{h}^{8}\), \(O_{h}^{10}\), 12 inequivalent atoms for \(O_{h}^{1}\), \(O_{h}^{2}\), \(O_{h}^{4}\), \(O_{h}^{5}\) and 24 inequivalent atoms for \(O_{h}^{7}\). This explains the absence of \(\Gamma_{2}^{+}\) phonons in common \(O_{h}\) space group crystals: diamond (\#227; 2(\(a\)))~\cite{warren}, niobium monoxide (\#221; 3(\(c\)), 3(\(d\))), fluorite (\#225; 1(\(a\)), 2(\(c\)))~\cite{verstraete}, caesium chloride (\#221; 1(\(a\)), 1(\(b\))), sodium chloride (\#225; 1(\(a\)), 1(\(b\)))~\cite{burstein}, Pt\(_3\)O\(_4\) (\#229; 3(\(b\)), 4(\(c\))), cubic spinel (\#227; 2(\(a\)), 4(\(d\)), 8(\(e\)))~\cite{dewijs}, perovskite (\#221; 1(\(a\)), 3(\(c\)))~\cite{stirling} and others. Among all the \(O_{h}\) crystals that the authors are familiar with, \(A15\) is the only structure to have \(\Gamma_{2}^{+}\) phonon.

\begin{figure}
\begin{center}
\includegraphics[width=10cm]{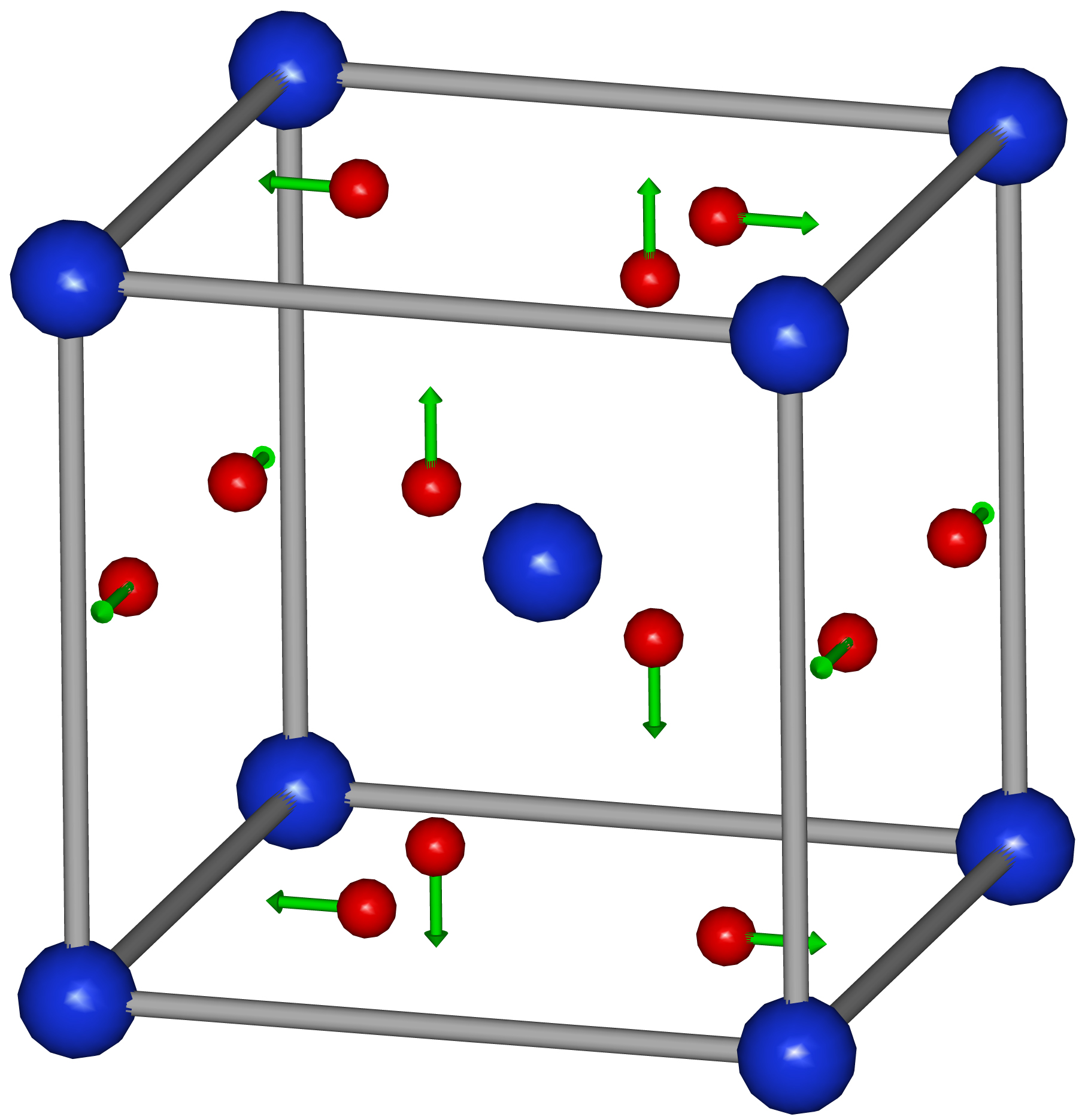}
\caption{The unit cell of Mo\(_3\)Si, a typical crystal of \(A15\) structure. Mo and Si atoms are indicated by small (blue) and big (red) spheres. The \(\Gamma_{2}^{+}\) phonon mode are shown by (red) arrows.} \label{fig:A15}
\end{center}
\end{figure}

The \(A15\) structure has simple cubic lattice. The general form for \(A15\) is A\(_3\)B. The primitive cell contains two formula units and the Wyckoff positions are 2(\(a\)) and 6(\(c\)) (see figure~\ref{fig:A15}). The required \(\Gamma_{2}^{+}\) phonon mode is shown in figure~\ref{fig:A15}. Based on discussions in section IV, one \(\Gamma_{3}^{+}\), one \(\Gamma_{4}^{+}\) and one \(\Gamma_{5}^{+}\) accompany the \(\Gamma_{2}^{+}\) phonon and the rest are three dimensional phonons. This prediction agrees with actual phonon calculations that the zone center phonons for \(A15\) crystals are \(\Gamma_{2}^{+}\) + \(\Gamma_{3}^{+}\) + \(\Gamma_{4}^{+}\) + \(\Gamma_{5}^{+}\) + 3 \(\Gamma_{4}^{-}\) + 2 \(\Gamma_{5}^{-}\)~\cite{Tutuncu}.  It is our next objective to perform the novel Raman scattering experiments on single crystals with \(A15\) structure, where \(\Gamma_{2}^{+}\) phonon of energy \(\sim\) 268 cm\(^{-1}\) can be well separated from its nearest neighbor of \(\Gamma_{3}^{+}\) phonon of energy \(\sim\) 284 cm\(^{-1}\) in case of Mo\(_3\)Si~\cite{Tutuncu}.

Compared with lattice systems of lower symmetry, the cubic system is perculiar in that the vector representation (\(x, y, z\)) belongs to a single three dimensional representation. This is the reason that one and two dimensional phonons happen rarely in crystals with cubic lattice. Table~\ref{tab:Oh} shows that some one dimensional phonons require as many as 24 atoms. Usually one does not go to such complicated strutures before trying simpler ones. On the other hand, although any set of Wyckoff positions with 48 inequivalent atoms, say, \(48(h)\) of \(O_{h}^{10}\) space group, gives all one dimensional phonons (\(\Gamma_{1}^{\pm}\), \(\Gamma_{2}^{\pm}\)), simpler crystal structures are better in that less phonon modes leads to larger seperations in energy and resolution is always an issue in spectroscopy experiments. The number of one dimensional phonons equals the number of two dimensional phonons therefore the searching for two dimensional phonon modes is as difficult as the searching for one dimensional phonon modes. This is also the case for magnetic space groups with cubic unitary groups. Following our tables, one can choose the crystal that is complicated enough to have the required (one or two dimensional) phonon mode, yet keeps the structure as simple as possible.\\

To summarize, four rules are obtained for phonon structure in \(T\), \(T_{h}\), \(T_{d}\), \(O\) and \(O_{h}\) space groups and for magnon structure with magnetic space groups whose unitary space part belongs to \(T\), \(T_{h}\), \(T_{d}\), \(O\) and \(O_{h}\) space groups:
\begin{enumerate}
\item At least four inequivalent atoms are needed to produce one dimensional phonon (magnon).
\item The number of one dimensional phonons (magnons) equals the number of two dimensional phonons (magnons) therefore at least four inequivalent atoms are needed to produce two dimensional phonons (magnons).
\item Three dimensional phonons (magnons) always exist, no matter how simple the crystal is.
\item A restriction rule is obtained:  \(n \times 3 \geq {\sum_{\gamma}{m \times 12}}\), \(n\) being the number of atoms in one Wyckoff set and \(m\) the number of one dimensional phonons (magnons).
\end{enumerate}

\end{document}